# Structural and flow approaches to complex network systems lesions scale analysis

Olexandr Polishchuk

Laboratory of Modeling and Optimization of Complex Systems
Pidstryhach Institute for Applied Problems of Mechanics and Mathematics, National Academy of Sciences of Ukraine, Lviv, Ukraine
od_polishchuk@ukr.net

**Abstract.** A comparative analysis of structural and flow approaches to analysis of vulnerability of complex network systems (NS) from targeted attacks and non-target lesions of various types was carried out. Typical structural and functional scenarios of successive targeted attacks on the most important by certain characteristics system elements were considered, and scenarios of simultaneous group attacks on the most significant NS's components were proposed. The problem of system lesions scale from heterogeneous negative impacts was investigated and it was confirmed that the flow approach allows us to obtain a much more realistic picture of consequences of such lesions. It is shown that scenarios of group targeted attacks built on the basis of the NC flow model are more optimal from the point of view of choosing attack targets than structural ones.

## 1. Вступ

За останні 5 років людство зіштовхнулося з двома глобальними викликами – пандемією Covid-19 та російсько-українською війною і викликаними нею всеосяжними санкціями до країни-агресора. Ці події є яскравими прикладами цілеспрямованих атак та нецільового ураження практично усіх сфер людської життєдіяльності [1, 2]. Ураження реальних систем можуть бути локальними, груповими та загальносистемними, причому локальні можуть переростати у групові, а групові – у загальносистемні [3]. Вони можуть бути передбачуваними та неочікуваними, централізованими та децентралізованими, коли уражений елемент сам стає джерелом ураження, поширюватися з різною швидкістю у просторі і часі – від майже миттєвих (каскадні явища) до тих, які тривають десятиліттями (глобальне потепління, поширення СНІДу) [4, 5] і т. ін. Незважаючи на різнотипність реальних систем та розмаїття видів негативних впливів на них, цілеспрямовані атаки та нецільові ураження можуть мати багато спільного як у способах дії, так і наслідках таких впливів: поширення небезпечних інфекційних захворювань та комп'ютерних вірусів, автомобільні затори та *DDoS*-атаки, обстріли населених пунктів та землетруси [6] тощо.

Натепер основна увага дослідників у теорії складних мереж (ТСМ) зосереджена на розробленні стратегій захисту від послідовних цілеспрямованих атак на найважливіші зі структурного погляду вузли мережевих систем [7, 8]. Створенню сценаріїв атак на процес функціонування таких систем приділяється значно менше уваги. Безумовно, ураження структури впливає на цей процес, але дестабілізувати або навіть припинити функціонування системи можна і за неураженої структури (під час російсько-української війни авіатранспортна система України повністю припинила свою діяльність лише із-за загрози збиття літаків) [9]. Ще одним недоліком структурного підходу до аналізу уразливості МС є оцінка масштабності ураження [10]. Безпосередньо ураженими вважаються фактично знищені елементи, які необхідно вилучити зі структури системи. Опосередковано постраждалими обґрунтовано можна вважати лише суміжні до безпосередньо уражених вузли мережевої системи. Такий підхід є цілком прийнятний для асортативних мереж [11]. Однак, для дисасортативних МС, до яких відносяться більшість створених людиною систем та елементи яких пов'язані шляхами, такий підхід не відображає у повній мірі усю сукупність постраждалих унаслідок цілеспрямованої атаки або нецільового ураження елементів системи. Саме об'єктивна оцінка реальної картини наслідків негативного впливу, яка поєднує не лише безпосередньо уражені, але й усі опосередковано постраждалі елементи системи надасть змогу точніше класифікувати тип цього впливу та кількісно визначити заподіяну ним шкоду.

Одним із способів захисту системи є нейтралізація джерела негативного впливу, зокрема, контратака на нього (санкції проти країни-агресора, знищення сільськогосподарських шкідників, припинення діяльності терористичних та хакерських груп, вакцинація [12, 13] і т. ін.). Проблемі оптимізації сценаріїв таких атак, як і розробленню ефективних засобів протидії нецільовим ураженням натепер також приділяють достатньо мало уваги, хоча сторона, яка ініціює та здійснює таку контратаку, наприклад, вводить санкції проти країни-агресора або карантинні заходи під час епідемій небезпечних інфекційних захворювань також несе чималі втрати. Для принаймні часткового вирішення перерахованих вище проблем у статті пропонується потоковий підхід до аналізу уразливості процесу функціонування МС та показуються його переваги над структурним підходом під час оцінювання реальних втрат, заподіяних негативним впливом на систему, а також оптимізації сценаріїв цілеспрямованих атак на неї.

## 2. Атаки на структуру мережевої системи

Математичною моделлю складної мережі (СМ) $G = (V, E)$, де $V$ – множина вузлів мережі $G$ та $E$ – множина зв'язків між ними, є матриця суміжності $\mathbf{A} = \{a_{ij}\}_{i,j=1}^{N}$, $N$ – кількість вузлів, які входять до складу мережі [14]. Значення $a_{ij}$ матриці $\mathbf{A}$ дорівнює 1, якщо зв'язок між вузлами $n_i$ та $n_j$ існує (такі вузли називаються суміжними), та дорівнює 0, якщо такий зв'язок відсутній. Також припускається, що вузли не мають зв'язків-петель, тобто діагональні елементи матриці $\mathbf{A}$ є нульовими.

Жодна велика складна система не може захистити усі свої елементи [15, 16]. Тому постає проблема визначення тих вузлів МС, які необхідно захистити насамперед. Для вирішення цієї проблеми у ТСМ у якості показника важливості вузла використовується поняття центральності. Основними локальними характеристиками вузла $n_i$ бінарної орієнтованої мережі є його вхідний та вихідний ступені [14] або центральності за ступенем. Тут під локальною ми розуміємо характеристику, яка описує властивості самого елемента або той чи інший аспект його взаємодії із суміжними елементами системи. Вхідний ступінь вузла визначає кількість зв'язків, які «входять» до нього, а вихідний ступінь – кількість зв'язків, які «виходять» із даного вузла до суміжних вузлів МС. Глобальними називатимемо характеристики елемента, які описують той або інший аспект його взаємодії з усіма іншими елементами цієї системи. Глобальна центральність дозволяє визначати важливість вузла у мережі загалом. Однак, поняття «важливість» може мати різний зміст, що призвело до появи багатьох визначень терміну «центральність», а саме посередництва, близькості, по власному вектору, Фрімана [3, 17] тощо (всього їх налічується близько 20). При цьому значення однієї центральності може суперечити іншому та центральність, важлива для вирішення однієї проблеми, може бути несуттєвою для іншої [18], що вносить певну неоднозначність під час їх використання у якості показника важливості елемента в структурі системи.

Зазвичай під атакою у ТСМ розуміються дії, спрямовані на умисне видалення зі структури системи певної кількості найважливіших за обраною центральністю вузлів з метою зміни структурних властивостей мережі [8, 19]. Оскільки ураження системи зазвичай здійснюється шляхом послідовного або одночасного ураження її елементів, то першим кроком під час формування способів її захисту є побудова так званих сценаріїв цілеспрямованих атак на мережеву систему [20, 21]. Найбільш дієві сценарії таких атак формуються, коли їх розробник «ставить себе» на місце «зловмисника», який мінімальними засобами намагається завдати максимальної шкоди атакованій системі. Розробленню кожного сценарію повинно передувати вироблення критеріїв успішності атаки. Зі структурного погляду такими критеріями можуть бути поділ СМ на незв'язні складові, збільшення середньої довжини найкоротшого шляху [19, 22] і т. ін.

Розроблені натепер структурні сценарії ураження МС, які можна поділити на дві основні групи, базуються на використанні згаданих вище центральностей вузлів у структурі системи (узагальненої центральності за ступенем, як суми вхідного та вихідного ступеня вузла, центральностей за посередництвом, близькістю, власним значенням тощо) [23, 24]. Кожний із сценаріїв першої групи починається із впорядкування множини вузлів МС згідно зменшення значень їх центральності обраного типу та подальшому послідовному вилученні зі структури вузлів згідно цього порядку поки не буде виконаний критерій успішності атаки. Сценарії цієї групи не передбачають зміну значень центральностей вузлів, які залишились у мережі. У другій групі сценаріїв враховується, що з кожним вилученням вузла структура МС змінюється завдяки налагодженню нових зв'язків між вузлами, що залишились. Це потребує нового впорядкування послідовності вузлів МС згідно змінених значень їх центральності обраного типу. На наступному кроці цієї групи сценаріїв вилучається вузол з початку новоствореного списку, який враховує ці зміни. Очевидно, що описані вище типові сценарії не визначають конкретні способи захисту реальної системи, які залежать від її виду та типу негативного впливу, однак, ці сценарії дають змогу ідентифікувати елементи, які необхідно першочергово захистити з погляду їх важливості у структурі МС.

Загалом для захисту мережевих систем від різнорідних негативних впливів постають три види взаємопов'язаних завдань, а саме: аналіз реальних та потенційних загроз та розроблення засобів ефективного захисту від них *до* ураження; забезпечення способів протидії поширенню негативних впливів і мінімізації їх наслідків *під час* ураження та оцінювання наслідків і відновлення структури та процесу функціонування системи *після* її ураження. Адже, чим краще захищена система, тим слабшою є дія негативного впливу, а отже меншими його наслідки. Для вирішення першого із цих завдань можна застосовувати наведені вище типові сценарії цілеспрямованих атак, а для другого та третього – структурну модель мережевої системи. Так, різниця рангу матриці **A** до ураження та рангу цієї матриці під час (після) ураження визначає кількість знищених під час (після) нього вузлів у вихідній структурі МС. Різниця кількості ненульових елементів структурної моделі мережевої системи до ураження та кількості ненульових елементів матриці **A** під час (після) ураження визначає кількість знищених під час (після) нього ребер у вихідній структурі МС. Таким чином, порівняння структурних моделей мережевої системи дає змогу скласти достатньо об'єктивне кількісне уявлення про рівень ураження системи або окремих її складових унаслідок здійсненої цілеспрямованої атаки або дії нецільового ураження. Водночас, поряд із інтегральними показниками ураження (сумарною кількістю знищених вузлів та ребер мережі), структурна модель дає змогу аналізувати ураження кожного елемента МС. Так, обнулення елемента матриці **A** свідчить про вилучення (знищення, блокування) відповідного ребра зі складу структури мережевої системи, обнулення всіх елементів рядка та стовпця матриці **A** – про вилучення відповідного вузла зі складу структури МС, зменшення узагальненого ступеня вузла – про скорочення кількості його взаємодій з іншими елементами системи. Загалом, структурна модель дає змогу відтворити картину усіх безпосередньо уражених та частини опосередковано постраждалих елементів МС. Відношення кількості реально уражених до кількості атакованих елементів системи є об'єктивним показником її захищеності від негативного впливу певного виду. Основним недоліком структурного підходу для оцінки наслідків ураження є те, що опосередковано постраждалими можна вважати лише суміжні до безпосередньо уражених вузли мережевої системи (рис. 1, чорним позначені безпосередньо уражені вузли МС; сірим – суміжні із безпосередньо ураженими (опосередковано постраждалі) вузли; білим – неуражені вузли мережевої системи; неперервна крива обмежує безпосередньо уражену область МС; штрихова крива – суміжну із безпосередньо ураженою, тобто область опосередковано постраждалих елементів мережевої системи).

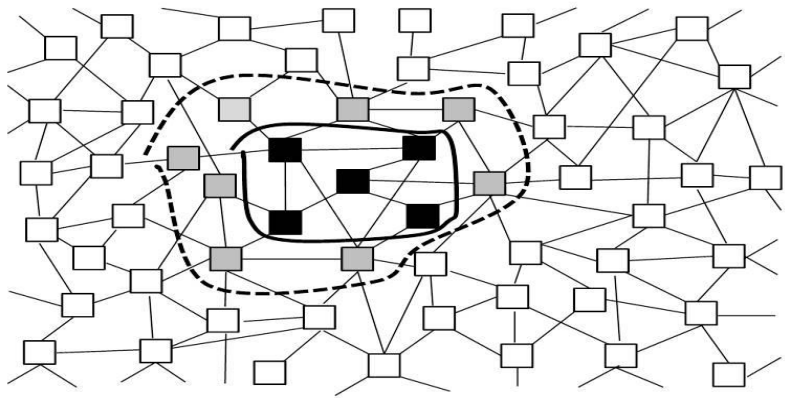

Рис. 1. Оцінка наслідків цілеспрямованої атаки на підставі структурної моделі МС

Очевидно, що одночасні групові атаки чи нецільові ураження системи є значно складнішими за послідовні поелементні як з погляду її захисту, так і подолання наслідків. Ми поділяємо одночасні групові негативні впливи на одноразові (атака Аль-Каїди на США 11 вересня 2001 року), повторні (18 ракетних ударів по Києву у травні 2023 року) та послідовні (напади на трансформаторні станції енергосистеми України у 2022 – 2024 рр.). Повторні групові атаки здійснюються регулярно через певні проміжки часу на одні і ті ж об'єкти системи. Послідовні групові атаки відрізняються від повторних зміною цілей ураження. Особлива небезпека полягає у тому, що успішні послідовні групові атаки можуть призвести до загальносистемного ураження МС, наприклад, тривалого блекауту у країні. У випадку цілеспрямованих атак цей поділ часто визначається спроможністю нападника здійснити наступні масовані напади та здатністю атакованої системи ефективно захищатися та протидіяти їм. Зрозуміло, що кожний із перерахованих вище видів атаки потребує розроблення специфічного типу сценаріїв її найвірогіднішої реалізації. Найпростіший сценарій одноразової групової атаки очевидно реалізується спробою одномоментного ураження групи найважливіших за визначеною центральністю елементів МС. Сценарій повторної атаки реалізується спробою ураження попередньо обраної та раніше атакованої, але не знищеної групи елементів мережевої системи. Сценарій послідовної групової атаки передбачає поступове виконання наступних кроків:
1) складаємо перелік груп вузлів (підсистем) МС у порядку зменшення обраних за певною ознакою показників їх важливості у системі;
2) видаляємо першу групу із створеного переліку;
3) якщо критерій успішності атаки досягнуто, то завершуємо виконання сценарію, інакше переходимо до пункту 4;
4) оскільки структура та процес функціонування системи унаслідок видалення певної групи вузлів (та їх зв'язків) змінюється, складаємо новий перелік груп у порядку зменшення переобчислених показників їх важливості у МС та переходимо до пункту 2.

Якщо під час реалізації крайнього сценарію певна група вузлів містить занадто велику кількість елементів, які нападник неспроможний уразити одночасно, то таку групу поділяємо на мінімальне число зв'язних підгруп, доступних для таких атак (російський агресор протягом нападу на Україну випускав по 100-150 ракет та БПЛА одночасно, але не по 500 із-за браку відповідних ресурсів). Окрім того, виконання сценарію може завершитись, коли у атакуючої сторони вичерпані ресурси для продовження атаки. Із наведених вище сценаріїв слідує, що основним способом підвищення їх ефективності є вибір показників важливості групи в системі, ураження якої завдасть їй якнайбільшої шкоди [25]. Найбільш очевидний спосіб такого вибору полягає у формуванні переліку вузлів МС у порядку зменшення значень їх центральності обраного типу та формуванні групи із перших вузлів цього переліку, кількість яких визначається спроможністю нападника здійснити одночасну атаку на них. Другий спосіб базується на принципі ієрархії вкладеності мережевої системи [26]. Пропонований нами спосіб полягає у застосуванні поняття $k$-серцевини СМ, як найбільшої підмережі вихідної мережі, центральність за узагальненим структурним ступенем вузлів якої є не меншим значення $k>1$ [27]. Цей спосіб базується на використанні найважливіших зі структурного погляду складових мережі та очевидним чином вкладається у наведений вище сценарій послідовних групових атак. Зокрема, спочатку групи виділяються для максимального для даної СМ значення $k$, яке потім послідовно зменшується до виконання критерію успішності атаки.

### 3. Атаки на процес функціонування мережевої системи

Для визначення функціональних показників важливості окремих складових МС скористаємося її потоковою моделлю [3]. Під потоком, який проходить ребром мережі, ми розуміємо певну, співвіднесену до цього ребра, додатну функцію. Ця функція може відображати щільність потоку у кожній точці ребра або об'єм потоку, який знаходиться на ребрі у поточний момент часу $t \geq 0$, чи сумарний об'єм потоку, який пройшов ребром мережі до поточного моменту за певний період тривалістю $T>0$ і т. ін. Відобразимо сукупність потоків, які проходять ребрами МС, у вигляді потокової матриці суміжності $\mathbf{V}(t)$, елементи якої визначаються об'ємами потоків, які пройшли ребрами складної мережі $G$ за період $[t-T, t]$ до поточного моменту часу $t \geq T$:

$$\mathbf{V}(t) = \{V_{ij}(t)\}_{i,j=1}^{N}, \quad V_{ij}(t) = \tilde{V}_{ij}(t) \Big/ \max_{l,m=\overline{1,N}} \tilde{V}_{lm}(t), \quad i,j = \overline{1,N},$$

у якій значення $\tilde{V}_{ij}(t)$ дорівнюють реальним обсягам потоків, які пройшли ребром $(n_i, n_j)$, $i,j = \overline{1,N}$, складної мережі за проміжок часу $[t-T, t]$, $t \geq T$. Очевидно, що структура матриці $\mathbf{V}(t)$ співпадає зі структурою матриці $\mathbf{A}$. Елементи потокової матриці суміжності МС визначаються на підставі емпіричних даних про рух потоків її ребрами. Натепер за допомогою сучасних засобів відбору інформації такі дані достатньо легко отримати для багатьох природних та переважної більшості створених людиною систем (транспортних, енергетичних, фінансових, інформаційних тощо) [28]. Зрозуміло, що описана вище потокова модель МС не є її математичною моделлю у звичному розумінні цього слова, але вона дає достатньо чітке кількісне уявлення про процес функціонування мережевої системи, дозволяє аналізувати особливості та прогнозувати поведінку цих процесів, а також оцінювати їх ефективність та запобігати існуючим або потенційним загрозам [3].

Інший порівняно зі структурним, і часто значно дієвіший та простіший для реалізації спосіб атаки полягає у дестабілізації або припиненні процесу функціонування окремих складових або системи загалом без безпосереднього знищення її елементів – суттєвого скорочення або припинення руху потоків мережею, створенні умов критичного завантаження шляхів руху цих потоків, блокуванні окремих вузлів-генераторів, транзитерів та/або кінцевих приймачів потоків, десинхронізації руху потоків мережею тощо. Побудова сценаріїв послідовних цілеспрямованих атак на найважливіші з функціонального погляду елементи системи здійснюється за тими ж принципами, що й типових структурних, з тією різницею, що у якості показників важливості вузлів МС використовуються характеристики, які визначають роль елементів МС у процесі її функціонування як генераторів, кінцевих приймачів та транзитерів потоків [3]. Структурний та функціональний підходи до побудови сценаріїв цілеспрямованих атак на систему можна поєднувати. Наприклад, якщо у послідовності вузлів МС існують групи з однаковими значеннями певного типу функціональної центральності, їх можна впорядкувати за значеннями обраного типу структурної центральності і навпаки.

У якості інтегрального показника втрат, заподіяних мережевій системі певним негативним впливом, можна використовувати різницю суми елементів матриці $\mathbf{V}(t)$ до ураження та суми елементів цієї матриці під час (після) ураження. Цей показник визначає сумарне зменшення обсягів потоків, в системі унаслідок дії негативного впливу. Водночас, поряд із інтегральними показниками, потокова модель дає змогу аналізувати ураження кожного елемента МС. Так, обнулення елемента матриці $\mathbf{V}(t)$ свідчить про вилучення (знищення, блокування) відповідного ребра з процесу функціонування мережевої системи, обнулення всіх елементів рядка та стовпця матриці $\mathbf{V}(t)$, які відповідають певному вузлу МС, – про вилучення цього вузла з процесу функціонування системи. Очевидно, що саме такі елементи МС визначаються за допомогою структурної моделі мережевої системи. Зменшення значення елемента матриці $\mathbf{V}(t)$ є ознакою зменшення обсягів потоків, які проходять відповідним ребром, а зменшення значення суми елементів у рядку та стовпці, які відповідають певному вузлу мережевої системи – про зменшення обсягів потоків, які генеруються, приймаються та проходять транзитом через цей вузол МС. Загалом, потокова модель дає змогу відтворити картину не тільки знищених, але й усіх опосередковано постраждалих вузлів та ребер мережевої системи та кількісно визначати рівень заподіяної шкоди, що є додатковою перевагою цієї моделі. На рис. 2 зображені наслідки цілеспрямованої атаки на мережеву систему, отримані на підставі її потокової моделі (чорним позначені безпосередньо уражені вузли МС; сірим – вузли, обсяги руху потоків з (до, через) яких зменшились унаслідок ураження; білим – неуражені вузли мережевої системи; неперервна крива обмежує безпосередньо уражену область МС; штрихова крива – суміжну із безпосередньо ураженою область мережевої системи; точкова крива – опосередковано постраждалу область МС). Як слідує із цього рисунка, визначена на підставі потокової моделі область опосередковано постраждалих елементів МС може бути значно більшою, ніж визначена на підставі структурної моделі область суміжних із безпосередньо ураженими вузлами мережевої системи. Таким чином, порівняння потокових моделей МС до, під час та після негативного впливу дає змогу скласти достатньо об'єктивне кількісне уявлення про рівень ураження мережевої системи або окремих її складових унаслідок здійсненої цілеспрямованої атаки або дії нецільового ураження. Відношення кількості реально уражених до кількості атакованих елементів системи є об'єктивним показником її захищеності від атак певного виду.

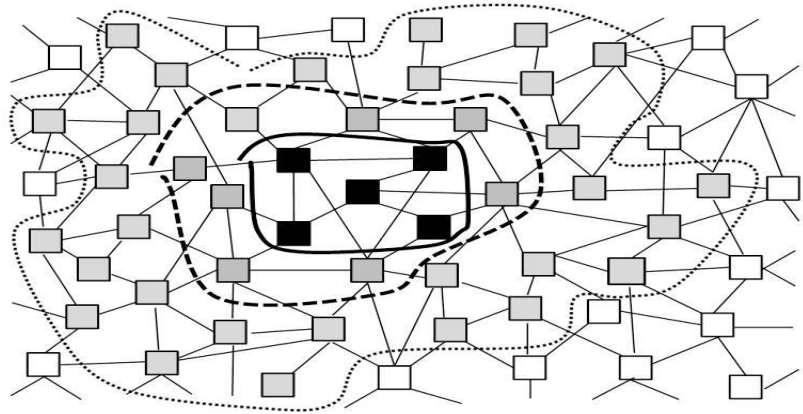

Рис. 2. Оцінка наслідків цілеспрямованої атаки на підставі потокової моделі МС

На підставі потокової моделі ми можемо визначити такі глобальні характеристики вузлів МС, як вхідні та вихідні параметри їхнього впливу на систему, а також параметри посередництва [3]. А саме, вхідною (вихідною) силою впливу вузла – кінцевого приймача (генератора) потоків вважатимемо сумарні обсяги потоків, які були прийняті (згенеровані) у цьому вузлі за період $[t-T,t]$; вхідною (вихідною) областю впливу вузла – кінцевого приймача (генератора) потоків вважатимемо сукупність вузлів МС, у яких були згенеровані (кінцево прийняті) спрямовані до (з) нього потоки за період $[t-T,t]$; вхідна (вихідна) потужність впливу вузла – кінцевого приймача (генератора) потоків дорівнює кількості елементів областей вхідного (вихідного) впливу цього вузла відповідно. Мірою посередництва вузла вважатимемо сумарні обсяги потоків, які пройшли через нього транзитом за проміжок часу $[t-T,t]$, $t \geq T$; областю посередництва вузла називатимемо сукупність вузлів МС, які спрямовували та з яких приймались потоки через даний транзитний вузол, а потужністю посередництва – кількість елементів, які входять до складу області посередництва. Загалом, після ураження певного вузла МС поєднання областей його впливу та посередництва повністю визначає сукупність та кількість усіх опосередковано постраждалих унаслідок цього елементів системи.

На рис. 3 зображені наслідки цілеспрямованої атаки на мережеву систему на підставі аналізу поведінки параметрів впливу та посередництва елементів мережевої системи (чорним позначені безпосередньо уражені вузли МС; сірими квадратами – суміжні із безпосередньо ураженими вузли; сірими ромбами, трикутниками та кругами – опосередковано постраждалі вузли-генератори, кінцеві приймачі та транзитери потоків відповідно; білим – неуражені вузли МС; неперервна крива обмежує безпосередньо уражену область МС; штрихова крива – суміжну із безпосередньо ураженою область мережевої системи; точкова крива – опосередковано постраждалу область МС).

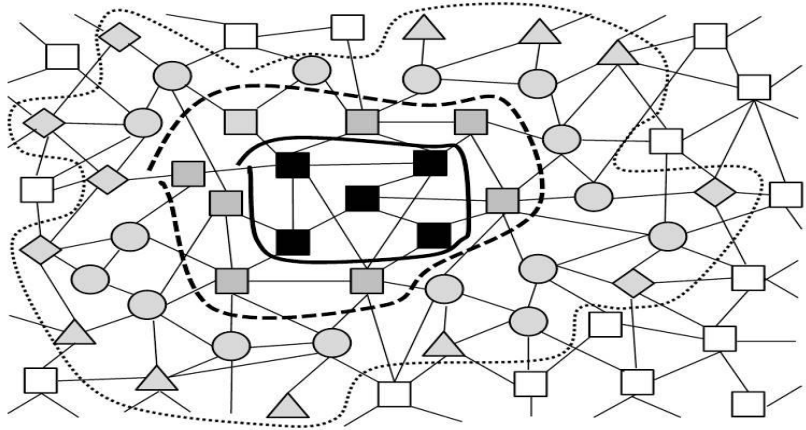

Рис. 3. Оцінка наслідків цілеспрямованої атаки на підставі аналізу параметрів впливу та посередництва елементів мережевої системи

Порівнюючи рис. 1, 2 та 3, можна зробити обґрунтований висновок, що потоковий підхід дає змогу створювати набагато реалістичнішу картину наслідків ураження, спричиненого певним негативним впливом, ніж структурний. Важливість аналізу ураження вузлів-генераторів, кінцевих приймачів та транзитерів потоків пояснюється тим, що вони потребують пошуку нових постачальників, споживачів та альтернативних шляхів руху потоків, що зазвичай є достатньо складною проблемою, особливо у випадку масових уражень МС. Під час розроблення сценаріїв одночасних групових атак у якості функціонально найважливішої складової мережевої системи можна застосовувати поняття її потокової $\lambda$-серцевини [3], як найбільшої підсистеми вихідної системи, елементи потокової матриці суміжності якої мають значення не менші ніж $\lambda \in [0, 1]$. Очевидно, що чим більшим є значення $\lambda$, тим важливішу з функціонального погляду складову МС відображає її $\lambda$-серцевина. Саме вона може стати однією із першочергових цілей одночасної групової атаки, сценарій якої був наведений у попередній секції. Аналогічно, як для елементів МС, ми можемо визначити параметри впливу та посередництва її $\lambda$-серцевини, що суттєво поглиблює аналіз ураження мережевої системи.

### 4. Оптимізація сценаріїв цілеспрямованих атак

Вище зазначалося, що одним із способів захисту системи є контратака на нападника. У випадку російської агресії проти України таким способом є фінансово-економічні санкції, контрнаступи для звільнення захоплених агресором територій країни, знищення його бойових підрозділів, логістичних вузлів та центрів управління військами тощо. Зрозуміло, що організатори таких контратак також несуть чималі втрати. Тобто, постає проблема оптимізації сценаріїв атак, а саме, як знищивши або заблокувавши роботу мінімальної кількості вузлів атакованої системи, заподіяти їй якнайбільшої шкоди. Подібна ситуація спостерігається під час розроблення сценаріїв протидії поширенню нецільових уражень, наприклад, епідемій небезпечних інфекційних захворювань (Covid-19). Зокрема, яким чином, заблокувавши якнайменшу кількість вузлів, що забезпечують переміщення пасажиропотоків, мінімізувати об'єми руху цих потоків мережею. Очевидно, що при цьому доцільно врахувати не лише масштабність безпосереднього негативного впливу, але й масштабність опосередкованих наслідків ураження. Вище для побудови сценаріїв одночасних групових атак пропонувалось застосування понять структурної $k$- та потокової $\lambda$-серцевини МС. Покажемо, що використання потокових $\lambda$-серцевин порівняно зі структурними $k$-серцевинами мережевих систем є значно ефективнішим під час побудови сценаріїв цілеспрямованих атак як з погляду можливого ураження функціонально найважливіших елементів МС, так і з метою оптимізації цих сценаріїв за кількістю об'єктів атаки. Розглянемо залізничну транспортну систему (ЗТС) західного регіону України. На рис. 4а відображено структуру цієї системи, а на рис. 4б – цю ж структуру, але у вигляді зваженої мережі, яка схематично відображає об'єми руху вантажних потоків, які пройшли її ребрами протягом 2021 року (товщина ліній є пропорційною об'ємам потоків). Зазначимо, що загалом ця мережа містить 354 вузли, однак на рис. 4а-б відображено лише 29 вузлів та 62 ребра (транзитні вузли зі структурним ступенем 2 не відображаються, а ребром вважаємо лінію, яка поєднує два вузли зі ступенем, більшим ніж 2). На рис. 4в міститься структурна $4$-серцевина ЗТС, до складу якої входить 12 вузлів та 35 ребер, а на рис. 4г – її потокова $0,8$-серцевина, яка містить 4 вузла та 12 ребер. Одним із основних недоліків $k$-серцевин порівняно із потоковими серцевинами є можливість виключення функціонально важливих складових мережевої системи (шлях А-В на рис. 4г).

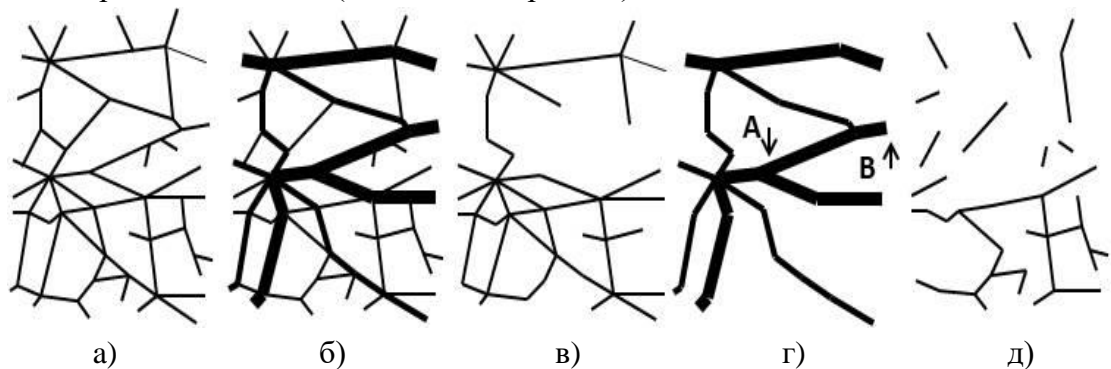

Рис. 4. Приклади структури (а), процесу функціонування (б), структурної $4$-серцевини (в), потокової $0,8$-серцевини (г) та доповнення до потокової $0,8$-серцевини у вихідній структурі (д) залізничної транспортної системи західного регіону України

Очевидно, що потокова $0,8$-серцевина відображає функціонально важливішу підсистему ЗТС і ціллю групової атаки на неї є значно менша кількість вузлів, ніж $4$-серцевини відповідної структури. Легко переконатись, що в обидвох випадках успішна атака на виділені за допомогою $k$- та $\lambda$-серцевин вузли МС призведе до фактичного припинення процесу її функціонування, оскільки поділяє її на незв'язні складові (рис. 4д), але в другому випадку мета атаки досягається значно меншими (у три рази з погляду кількості вузлів та ребер) зусиллями. Таким чином, потоковий підхід дає змогу будувати значно оптимальніші з погляду зусиль атакуючої сторони сценарії, ніж структурний. Шляхом аналізу параметрів впливу та посередництва $0,8$-серцевини наведеного фрагменту ЗТС було встановлено, що опосередковано постраждалими від успішної цілеспрямованої атаки на неї будуть усі елементи цього фрагменту.

## 5. Висновки

У 2019-2024 роках людство зіштовхнулося з двома глобальними викликами, перший з яких (пандемія Covid-19) є яскравим прикладом загальносистемного нецільового ураження, а другий – цілеспрямованої атаки (напад рф на Україну) та викликаної нею загрози світової продовольчої, енергетичної, фінансової кризи і зворотні всеосяжні санкції стосовно агресора, негативні наслідки яких торкнулися практично усіх країн світу. Людство виявилося непідготовленим до таких викликів, але на часі залишаються не менш небезпечні загрози. За минуле півстоліття зникло 67% відомих людині біологічних видів [29], а протягом останніх 20 років витрати на боротьбу з кліматичними катастрофами зросли у 8 разів [30]. Натепер ученим відомо більш ніж 20 вірусів небезпечних інфекційних захворювань, мутації яких можуть призвести до поширення пандемій, значно катастрофічніших за Covid-19 [31], посилюється загроза глобальних військових конфліктів тощо. Це підтверджує актуальність вивчення особливостей уражень складних мережевих систем та розроблення методів ефективного захисту від них. Розуміння структурної та функціональної важливості складових системи дає змогу обирати об'єкти, які потребують першочергового захисту або якнайшвидшого блокування, оскільки найбільш сприяють поширенню ураження. У роботі проведено порівняльний аналіз структурного та потокового підходів до аналізу уразливості складних мережевих систем від різнорідних негативних впливів, розглянуто типові сценарії послідовних цілеспрямованих атак на найважливіші за певними ознаками елементи системи та запропоновано сценарії одночасних групових атак на найважливіші складові МС. Досліджено проблему масштабності системних уражень від різнорідних негативних впливів та підтверджено, що потоковий підхід дає змогу отримати значно реалістичнішу картину наслідків таких уражень. Показано, що побудовані на підставі потокової моделі МС сценарії цілеспрямованих групових атак є оптимальнішими з погляду вибору цілей атаки, ніж структурні.